\documentclass[12pt]{article}

\usepackage{amsmath,amssymb}
\usepackage{tabularx}
\usepackage{epsfig}
\usepackage{graphicx}

\begin{document}

\begin{titlepage}

%\rightline{hep-th/yymmnnn}

%\vskip 2cm

\centerline{\large \bf {Supersymmetric dark matter, catalyzed BBN, and heavy moduli}}
\centerline{\large \bf {in mSUGRA with gravitino LSP and stau NLSP}}

\vskip 1cm

\centerline{Grigoris Panotopoulos}

\vskip 1cm

\centerline{ASC, Physics Department LMU,}

\vskip 0.2 cm

\centerline{Theresienstr. 37, 80333 Munich, Germany}

\vskip 0.2 cm

\centerline{email:{\it Grigoris.Panotopoulos@physik.uni-muenchen.de}}

\begin{abstract}
In mSUGRA model we assume that gravitino, the LSP, plays the role of cold dark matter in the universe, while the lightest stau, the NLSP, catalyzes primordial BBN reconciling the discrepancy between theory and observations. We have taken into account all gravitino production mechanisms, namely decay from heavy scalar fields, decay from the NLSP, and from the thermal bath. We find that the dark matter constraint is incompatible with the lower bound on the reheating temperature.
\end{abstract}

\end{titlepage}

There is accumulated evidence both from astrophysics and cosmology that about 1/4 of the energy budget of the universe consists of so called dark matter, namely a component which is non-relativistic and does not feel the electromagnetic nor the strong interaction. For a review on dark matter see e.g.~\cite{Munoz:2003gx}. Although the list of possible dark matter candidates is long, it is fair to say that the most popular dark matter particle is the LSP in supersymmetric models with R-parity conservation~\cite{Feng:2003zu}. The superpartners that have the right properties for playing the role of cold dark matter in the universe are the axino, the gravitino and the lightest neutralino. By far the most discussed case in the literature is the case of the neutralino (see the classical review~\cite{Jungman:1995df}), probably because of the prospects of possible detection.

On the other hand, primordial Big-Bang nucleosynthesis (BBN) is one of the cornerstones of modern cosmology. In the old days, BBN together with Hubble's law and CMB supported and strengthened  the Hot Big-Bang idea. Nowadays, BBN can be used to test and constrain possible new physics beyond the standard model. Although the general agreement between the theoretical predictions and the observed light nuclei abundances is quite impressive, it is true that that there is a discrepancy between the standard theory and Lithium isotopes observations~\cite{Cyburt:2006uv}. Recently it was proposed that the so called catalyzed BBN can reconcile the aforementioned discrepancy~\cite{Pospelov:2006sc}. Consider a heavy unstable negatively charged particle $X^-$ which can form Coulomb bound states together with Helium 4. Then the following reaction
\begin{equation}
(^4He X^-) + D \rightarrow ^6Li + X^-
\end{equation}
can affect the primordial light element abundances substantially. The Lithium 6 observations can constrain the properties of $X^-$, and in particular its lifetime~\cite{Pradler:2007is}.

In CMSSM the lightest neutralino or the lightest stau is typically the lightest superpartner. Therefore in scenarios in which gravitino is assumed to be the LSP, then the neutralino or the stau is the NSLP and therefore unstable with a lifetime which is typically larger than BBN time $t_{BBN} \sim 1$~sec. Energetic particles produced by the NLSP decay may dissociate the background nuclei and significantly affect the primordial abundances of the light elements. If such processes occur with sizable rates, the predictions of the standard BBN scenario are altered and the success of the primordial nucleosynthesis is spoiled. BBN constraints on cosmological scenarios with exotic long-lived particles predicted by physics beyond the Standard Model have been studied~\cite{constraints}, and the neutralino NLSP scenario is already disfavored~\cite{Feng:2004mt}. However the stau NLSP is still a viable scenario, and in this case the stau can play the role of the heavy unstable negatively charged particle $X^-$.

In the present article we work in the mSUGRA model and assume that the gravitino is the LSP while the stau is the NLSP. Gravitino plays the role of cold dark matter in the universe while the stau plays the role of $X^-$ to catalyze BBN. In the CMSSM, contrary to the MSSM, there is a small controllable number of parameters, namely four parameters and a sign. These are
\begin{itemize}
\item Universal gaugino masses
\begin{equation}
M_1(M_{GUT})=M_2(M_{GUT})=M_3(M_{GUT})=m_{1/2}
\end{equation}
\item Universal scalar masses
\begin{equation}
m_{\tilde{f}_i}(M_{GUT})=m_0
\end{equation}
\item Universal trilinear couplings
\begin{equation}
A_{i j}^u(M_{GUT}) = A_{i j}^d(M_{GUT}) = A_{i j}^l(M_{GUT}) = A_0 \delta_{i j}
\end{equation}
\item
\begin{equation}
tan \beta \equiv \frac{v_1}{v_2}
\end{equation}
where $v_1, v_2$ are the vevs of the Higgs doublets and $M_{GUT} \sim 10^{16}~GeV$ is the Grand Unification scale.
\end{itemize}
plus the sign of the $\mu$ parameter from the Higgs sector. In mSUGRA there is the additional condition for the gravitino mass, $m_{3/2}=m_0$~\cite{Ellis:2007mc}. The neutralino and stau masses depend on the universal scalar and gaugino mass as follows~\cite{Pradler:2007is,Cerdeno:2005eu}
\begin{eqnarray}
m_\chi & \simeq & 0.42 m_{1/2} \\
m_{\tilde{\tau}}^2 & \simeq & m_0^2+0.15 m_{1/2}^2
\end{eqnarray}
From the above formulas we can easily see the two limits in which either neutralino or stau is the lightest of the usual superpartners. In the limit in which $m_0 \gg m_{1/2}$, the stau mass $m_{\tilde{\tau}} \simeq m_0$ and in this case the neutralino is lighter than stau. In the limit in which $m_0 \ll m_{1/2}$, the stau mass $m_{\tilde{\tau}} \simeq 0.387 m_{1/2}$ and in this case the stau is lighter than the neutralino. The natural range for values of $m_{1/2}$ is from $100$~GeV up to a few TeV, and following~\cite{Pradler:2007is} we shall consider for $m_{1/2}$ the range
\begin{equation}
100~GeV \leq m_{1/2} \leq 6~TeV
\end{equation}
For the gravitino abundance we take all possible production mechanisms into account and impose the cold dark matter constraint~\cite{Spergel:2006hy}
\begin{equation}
0.075 < \Omega_{cdm} h^2=\Omega_{3/2} h^2 < 0.126
\end{equation}
At this point it is convenient to define the gravitino yield, $Y_{3/2} \equiv n_{3/2}/s$, where $n_{3/2}$ is the gravitino number density, $s=h_{eff}(T) \frac{2 \pi^2}{45} T^3$ is the entropy density for a relativistic thermal bath, and $h_{eff}$ counts the relativistic degrees of freedom. The gravitino abundance $\Omega_{3/2}$ in terms of the gravitino yield is given by
\begin{equation}
\Omega_{3/2} h^2=\frac{m_{3/2} s(T_0) Y_{3/2} h^2}{\rho_{cr}}=2.75 \times 10^{8} \left ( \frac{m_{3/2}}{GeV} \right ) Y_{3/2}(T_0)
\end{equation}
where we have used the values
\begin{eqnarray}
T_0 & = & 2.73 K=2.35 \times 10^{-13}~GeV \\
h_{eff}(T_0)& = & 3.91 \\
\rho_{cr}/h^2& = & 8.1 \times 10^{-47}~GeV^4
\end{eqnarray}
The total gravitino yield has three contributions, namely one from the thermal bath, one from the NLSP decay, and one more from heavy moduli decay.
\begin{equation}
Y_{3/2}=Y_{3/2}^{TP}+Y_{3/2}^{NLSP}+Y_{3/2}^{modulus}
\end{equation}
The contribution from the thermal production has been computed in~\cite{Bolz:2000fu,Pradler:2006qh,Rychkov:2007uq}. In~\cite{Bolz:2000fu} the gravitino production was computed in leading order in the gauge coupling $g_3$, in~\cite{Pradler:2006qh} the thermal rate was computed in leading order in all Standard Model gauge couplings $g_Y, g_2, g_3$, and in~\cite{Rychkov:2007uq} new effects were taken into account, namely: a) gravitino production via gluon $\rightarrow$ gluino $+$ gravitino and other decays, apart from the previously considered $2 \rightarrow 2$ gauge scatterings, b) the effect of the top Yukawa coupling, and c) a proper treatment of the reheating process. Here we shall use the result of~\cite{Bolz:2000fu} since the corrections of~\cite{Pradler:2006qh,Rychkov:2007uq} do not alter our conclusions. Therefore the thermal gravitino production is given by
\begin{equation}
Y_{3/2}^{TP}=1.1 \times 10^{-12} \: \left ( \frac{T_R}{10^{10}~GeV} \right ) \: \left ( \frac{m_{\tilde{g}}}{m_{3/2}} \right )^2
\end{equation}
The second contribution to the gravitino abundance comes from the decay of the NLSP
\begin{equation}
\Omega_{3/2}^{NLSP} h^2 = \frac{m_{3/2}}{m_{NLSP}} \: \Omega_{NLSP} h^2
\end{equation}
with $m_{3/2}$ the gravitino mass, $m_{NLSP}$ the mass of the NLSP (here the stau) and $\Omega_{NLSP} h^2$ the NLSP abundance had it did not decay into the gravitino, which for the stau is estimated to be~\cite{Kawasaki:2007xb}
\begin{equation}
\Omega_{\tilde{\tau}}h^2 \simeq \left( \frac{m_{\tilde{\tau}}}{2~TeV} \right )^2
\end{equation}
The decay width of stau to tau and gravitino is given by~\cite{Pradler:2007is,Feng:2004mt}
\begin{equation}
\frac{1}{\tau_{\tilde{\tau}}}=\Gamma(\tilde{\tau} \rightarrow \tau+\psi_{3/2})=\frac{1}{48 \pi M_p^2} \frac{m_{\tilde{\tau}}^5}{m_{3/2}^2} \left( 1-\frac{m_{3/2}^2}{m_{\tilde{\tau}}^2} \right)^4
\end{equation}
where $M_p=2.4 \times 10^{18}~GeV$ is the reduced Planck mass.
In the limit in which $m_{3/2} \ll m_{\tilde{\tau}}$ the stau lifetime is simplified as follows
\begin{equation}
\tau_{\tilde{\tau}}=6.1 \times 10^3~sec \left( \frac{m_{3/2}}{100~GeV} \right)^2 \left( \frac{1000~GeV}{m_{\tilde{\tau}}} \right)^5
\end{equation}
According to the results of~\cite{Pradler:2007is}, the catalyzed BBN imposes on the stau lifetime the upper bound
\begin{equation}
\tau_{\tilde{\tau}} \leq 5 \times 10^3~sec
\end{equation}
Using that $m_{3/2}=m_0$, $m_{\tilde{\tau}}=0.387 m_{1/2}$, and the expression for the stau lifetime we obtain
\begin{equation}
m_{1/2} \geq 426 \left( \frac{m_0}{GeV} \right)^{2/5}~GeV
\end{equation}
On the other hand, the catalyzed BBN only works if the lifetime of the $X^-$ particle is larger than a certain value
\begin{equation}
\tau_{\tilde{\tau}} > 10^3~sec
\end{equation}
or
\begin{equation}
m_{1/2} < 588 \left( \frac{m_0}{GeV} \right)^{2/5}~GeV
\end{equation}
For example, if $m_{3/2}=m_0=100~GeV$, then $m_{1/2}$ takes values within the interval
\begin{equation}
2.69~TeV \leq m_{1/2} < 3.71~TeV
\end{equation}
The last contribution to the gravitino abundance comes from a heavy modulus decay. Moduli are four-dimensional scalar fields with a potential, and they appear from higher dimensional supergravity/superstring theories upon compactification down to four dimensions. Moduli parameterize the shape and size of the manifold used for compactificationand, and typically their vaccum expectation value is of the order of the Planck mass $M_p$. Finally, generically and without fine-tuning their mass is close to the gravitino mass. For example in the KKLT model~\cite{KKLT} one finds that $m_T \simeq 4 \pi^2 m_{3/2}$, where $T$ is the total volume modulus. However, with fine tuning it is possible to have a heavy modulus and an arbitrarily light gravitino~\cite{KL}. The gravitino yield from modulus decay is given by~\cite{Nakamura:2006uc}
\begin{equation}
Y_{3/2}^{modulus}=\frac{3}{2} \: \frac{\Gamma_{3/2}}{\Gamma_{tot}} \: \frac{T_R}{m_X}
\end{equation}
where $T_R$ is the reheating temperature at the modulus decay, $m_X$ is the modulus mass, $\Gamma_{tot}$ is the total decay rate of the modulus, and $\Gamma_{3/2}$ is the modulus  decay rate to a pair of gravitinos. Using the supergravity lagrangian one can compute both
$\Gamma_{tot}$ and $\Gamma_{3/2}$. The modulus dominantly decays into gauge bosons and gauginos with a total decay width~\cite{Nakamura:2006uc}
\begin{equation}
\Gamma_{tot} \equiv \Gamma(X \rightarrow all) \simeq \Gamma(X \rightarrow gg)+\Gamma(X \rightarrow \tilde{g} \tilde{g})=\frac{3}{16 \pi} \frac{m_X^3}{M_p^2}
\end{equation}
while the decay width into a pair of gravitinos is given by~\cite{Nakamura:2006uc}
\begin{equation}
\Gamma_{3/2}=\frac{1}{288 \pi} \frac{m_X^3}{M_p^2}
\end{equation}
in the limit in which $m_X \gg m_{3/2}$. Therefore the branching ratio for the decay channel into a gravitino pair is given by
\begin{equation}
Br(X \rightarrow \psi_{3/2} \psi_{3/2})=\frac{\Gamma_{3/2}}{\Gamma_{tot}}=\frac{1}{54} \sim 0.01
\end{equation}
We assume that the modulus dominates the energy density when it decays and thus it reheats the universe. The reheating temperature at the modulus decay is determined as usual by the condition $H(T_R)=\Gamma_{tot}$, where $H(T)$ is the Hubble parameter as a function of the temperature during the radiation era. Thus one finds
\begin{equation}
T_R=\left ( \frac{90}{\pi^2 g_*(T_R)} \right )^{1/4} \sqrt{\Gamma_{tot} M_p}
\end{equation}
or
\begin{equation}
T_R=4.9 \times 10^{-3} \left ( \frac{10}{g_*(T_R)} \right )^{1/4} \left ( \frac{m_X}{10^5~GeV} \right )^{3/2}~GeV
\end{equation}
Notice that $T_R$ is determined entirely by the modulus mass. From BBN there is a lower bound on the reheating temperature, coming from the requirement that at BBN time all three neutrino species are thermalized~\cite{Kawasaki:2000en}. This limit reads
\begin{equation}
T_R \geq 7~MeV
\end{equation}
which in turn induces a lower bound on modulus mass
\begin{equation}
m_X \geq 1.5 \times 10^5~GeV
\end{equation}
For given values of gravitino mass and stau mass, the total gravitino abundance is a function of the modulus mass. In figure~1 we show the gravitino abundance as a function of modulus mass for $m_0=100$~GeV and $m_{1/2}=2.69$~TeV, while in figure~2 we show the gravitino abundance versus the modulus mass for $m_0=100$~GeV and $m_{1/2}=3.7$~TeV. In both figures the horizontal strip shows the allowed range for cold dark matter. The figures show that gravitino can explain the cold dark matter in the universe provided that the modulus mass is of the order of $GeV$, which is obviously in contradiction to the limit $m_X \geq 1.5 \times 10^5$~GeV. Apart from that, since the modulus field decays into two gravitinos, kinematics requires that $m_X > 200$~GeV. Therefore, for these two reasons we conclude that the scenario under investigation must be excluded. Finally, if we consider larger values of $m_{3/2}$ we find that the situation becomes even worse. The contribution from the NLSP decay increases and therefore the modulus mass must be lower than before.

We remark in passing that if the gravitino mass is treated as one extra parameter and can be low enough, then it possible to satisfy both the cold dark matter constraint and the lower bound on the modulus mass. For example, in figure~3 we show the gravitino abundance versus the modulus mass for a light gravitino, $m_{3/2}=0.1$~GeV, and $m_{1/2}=234$~GeV. We see that the gravitino abundance is within the allowed observational range provided that the modulus mass is $m_X \sim 10^6$~GeV, which satisfies the limit $m_X \geq 1.5 \times 10^5$~GeV. This implies that if the scenario under investigation is to be realized in nature, then the gravitino must be light or, put it differently, the correct supersymmetry breaking pattern might be the gauge mediated one~\cite{Giudice:1998bp}.

In summary, in the present work we have studied supersymmetric dark matter in mSUGRA assuming  that the gravitino is the LSP and that the stau is the NLSP. The gravitino plays the role of cold dark matter in the universe, while the stau catalyzes the standard BBN reconciling the discrepancy between theoretical predictions and observations on cosmic lithium isotopes observations. We have taken into account all three gravitino production mechanisms and have imposed the cold dark matter constraint. The production mechanisms for the gravitino are i) a heavy modulus decay, ii) the NLSP (stau) decay, and iii) scattering processes from the thermal bath. We have assumed that the modulus dominates the energy density when it decays and reheats the universe. The reheating temperature is determined entirely by the modulus mass. Constraints from BBN impose a lower bound on the reheating temperature and in turn a lower bound on the modulus mass, $m_X \geq 1.5 \times 10^5$~GeV. On the other hand, if we compute the total gravitino abundance and impose the cold dark matter constraint we find that the modulus mass should be much lower than the above lower limit. Therefore, we conclude that the scenario under investigation must be excluded.

\section*{Acknowledgments}

The present work was supported by project "Particle Cosmology".

\newpage

\begin{figure}
\centerline{\epsfig{figure=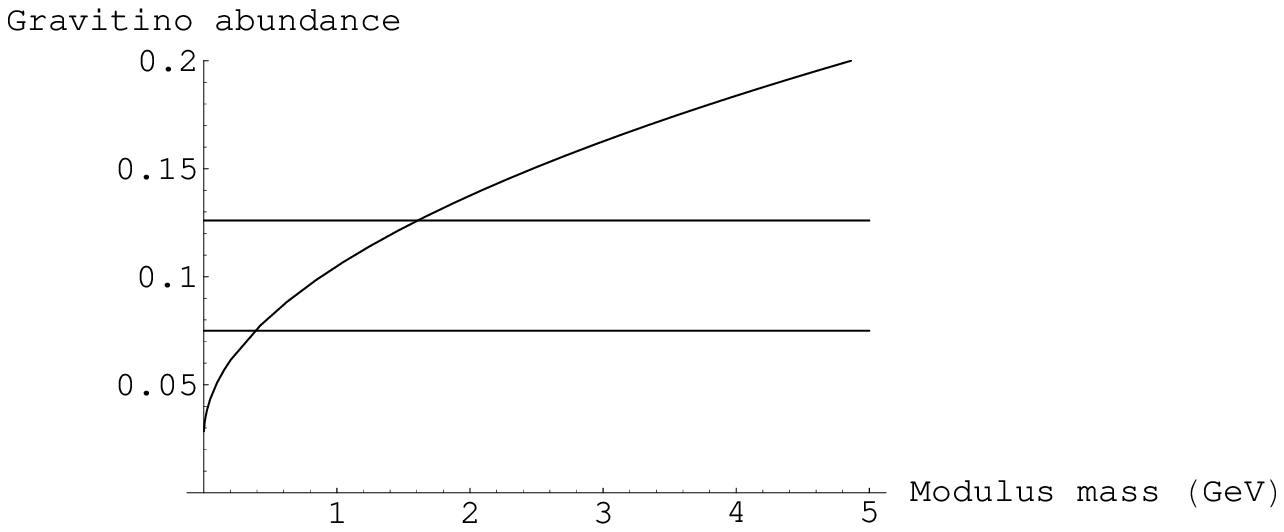,height=8cm,angle=0}}
\caption{Gravitino abundance versus modulus mass for $m_0=100$~GeV and $m_{1/2}=2.69$~TeV.
The strip shows the cold dark matter constraint.}
\end{figure}

\begin{figure}
\centerline{\epsfig{figure=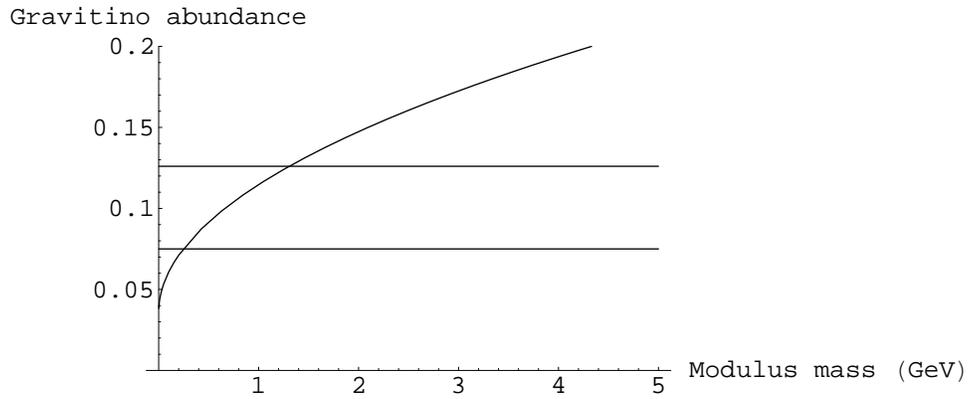,height=8cm,angle=0}}
\caption{Same as figure 1, but for $m_{1/2}=3.7$~TeV.}
\end{figure}

\begin{figure}
\centerline{\epsfig{figure=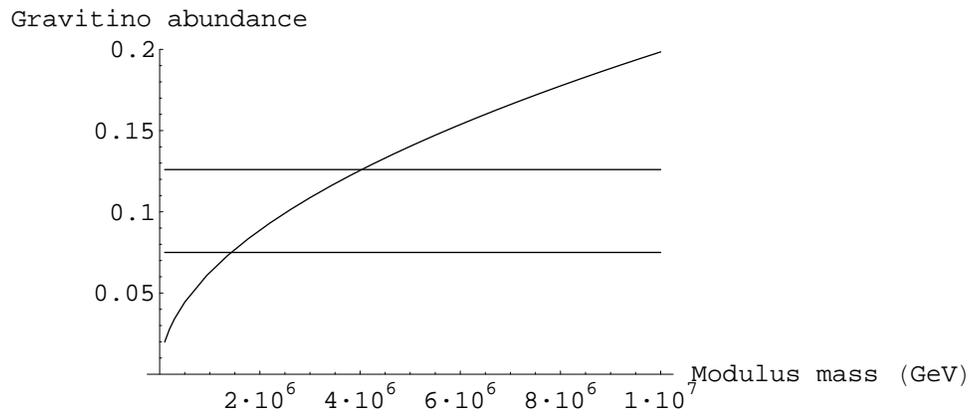,height=8cm,angle=0}}
\caption{Same as figure 1, but for $m_{3/2}=0.1$~GeV and $m_{1/2}=234$~GeV.}
\end{figure}

\end{document}